\documentclass[prl,aps,twocolumn,superscriptaddress]{revtex4-2}
\usepackage{amsfonts}

\usepackage{dcolumn}
\usepackage{bm}
\usepackage{tikz}
\usepackage[colorlinks=true,citecolor=blue,urlcolor=blue]{hyperref}
\usepackage{amsmath,amsfonts,amssymb,times,natbib}
\usepackage[standard]{ntheorem}

\begin{document}

\title{Excited-state spectroscopy of spin defects in hexagonal boron nitride}
\author
{Pei Yu$^{1,2,3\ast}$,
Haoyu Sun$^{1,2,3\ast}$,
Mengqi Wang$^{1,2,3}$,
Tao Zhang$^{1,2,3}$,
Xiangyu Ye$^{1,2,3}$,
Jingwei Zhou$^{1,2,3}$,
Hangyu Liu$^{1,2,3}$,
Chengjie Wang$^{1,2,3}$,
Fazhan Shi$^{1,2,3}$,
Ya Wang$^{1,2,3\dag}$,
Jiangfeng Du $^{1,2,3\dag}$
\\
\normalsize{$^{1}$ Hefei National Laboratory for Physical Sciences at the Microscale and School of Physical Sciences, University of Science and Technology of China, Hefei 230026, China.}\\
\normalsize{$^{2}$ CAS Key Laboratory of Microscale Magnetic Resonance, University of Science and Technology of China, Hefei 230026, China}\\
\normalsize{$^{3}$ CAS Center for Excellence in Quantum Information and Quantum Physics, University of Science and Technology of China, Hefei 230026, China}\\
\normalsize{$^{\ast}$ These authors contributed equally to this work.}\\
\normalsize{$^\dag$ Corresponding author. E-mail: ywustc@ustc.edu.cn, djf@ustc.edu.cn }
}

\begin{abstract}
We used optically detected magnetic resonance (ODMR) technique to directly probe electron-spin resonance transitions in the excited state of negatively-charged boron vacancy ($ V_B^- $) defects in hexagonal boron nitride (hBN) at room temperature. The data showed that the excited state has a zero-field splitting of $ \sim $ 2.1 GHz, a g-factor similar to the ground state and two types of hyperfine splitting $\sim$ 90 MHz and $\sim$ 18.8 MHz respectively. Pulsed ODMR experiments were conducted to further verify observed resonant peaks corresponding to spin transitions in the excited state. In addition, negative peaks in photoluminescence and ODMR contrast as a function of magnetic field magnitude and angle at level anti-crossing were observed and explained by coherent spin precession and anisotropic relaxation. This work provided significant insights for studying the structure of  $ V_B^- $ excited states, which might be used for quantum information processing and nanoscale quantum sensing.
\end{abstract}

\maketitle
The two-dimensional (2D) hexagonal boron nitride (hBN) hosting optically active defects has recently emerged as a promising system for quantum applications \cite{xia2014two,tran2016quantum}. Its 2D nature allows hBN transferred and integrated with solid devices like cavities and other heterogeneous materials in a controllable way so that the defects can have controllable interaction.\cite{tran2017deterministic,caldwell2019photonics,wang2017graphene}
Especially, for spin defects allowing optical initialization, readout and spin manipulation, the nanoscale proximity of the spin defects to target samples may allow for high-resolution quantum-sensing.
The potential spin defects include negatively-charged boron vacancy ($ V_B^- $) defects \cite{gottscholl2020initialization,chejanovsky2021single,stern2021room,gottscholl2021spin,liu2021temperature}, carbon-related defects\cite{stern2021room,chejanovsky2021single} and nitrogen-related defects\cite{chejanovsky2021single}.
For $ V_B^- $ defects, recent works have confirmed that it is a spin three-level system and revealed structure of the ground state (GS), which can be potentially used for sensing temperature, pressure, and magnetic field \cite{gottscholl2020initialization,froch2021coupling,gottscholl2021spin}. However, the excited state (ES) structure as well as the dynamics of excitation-emission cycles are not yet fully understood \cite{ivady2020ab,reimers2020photoluminescence}. Understanding this excited-state spin level is of critical importance for better control of $ V_B^- $ centers spin state and quantum sensing application.

Accessing the spin structure of the orbital ES has been an experimental challenge, partly because of the short lifetime. For the $ V_B^- $  color center, the excited state lifetime is only 1.2 ns at room temperature \cite{gottscholl2020initialization,mendelson2021coupling}. The photoluminescence spectrum of $ V_B^- $ is a broad band emission at 4K, centered at $ \approx $ 850 nm, which makes it difficult to study the energy level structure of the ES by resonant excitation \cite{reimers2020photoluminescence,mendelson2021coupling,batalov2009low}. In contrast, we present a simple approach that directly probes the ES spin transitions by using optically detected magnetic resonance (ODMR) technique at room temperature\cite{fuchs2008excited,neumann2009excited}. Through this technique, we obtained the detailed ES spin level structure including the hyperfine interaction to the surrounding nuclear spin, demonstrated controlled spin level mixing by means of magnetic field alignment, and showed the resulted tuning of optical illumination and ODMR contrast. These results deep our understanding of $ V_B^- $  color center and help for the next step of developing $ V_B^- $ color center spin as a solid-state quantum sensor.

The $ V_B^- $ center in the hexagonal boron nitride is a defect formed by the loss of a boron atom at its lattice site and it is surrounded by three equivalent nitrogen atoms. So far, this defect is considered to be a spin triplet in both the electronic GS($ {}^3A $) as well as the ES($ {}^3E $) \cite{gottscholl2020initialization,gottscholl2021room,ivady2020ab,reimers2020photoluminescence}. Other energy levels are still under debate, but it is now well established through experiments and theoretical calculations that at least one metastable state $ {}^3A $ is located between the ground state and the excited state, and long lifetime of the metastable state leads to low quantum yield, $ \sim $ 0.03$ \% $ \cite{ivady2020ab,reimers2020photoluminescence}. Similar to the nitrogen vacancy color center ($ NV^- $) in diamond, due to the presence of non-radiative transitions and spin selective relaxation through an intersystem crossing to $ {}^1A $, optical excitation-emission cycles induce the polarization of electron spin into the $ m_S = 0 $ spin state \cite{gottscholl2020initialization,jelezko2002single,harrison2004optical}. Furthermore, the spin selective nature of this process also creates a difference in the photoluminescence signal that depends on the spin state. In particular, the photoluminescence intensity is higher at $ m_S = 0 $  than at $ m_S = \pm1 $ spin state \cite{gottscholl2020initialization}.

\begin{figure}
	\centering
	{\includegraphics[width=1.0\columnwidth]{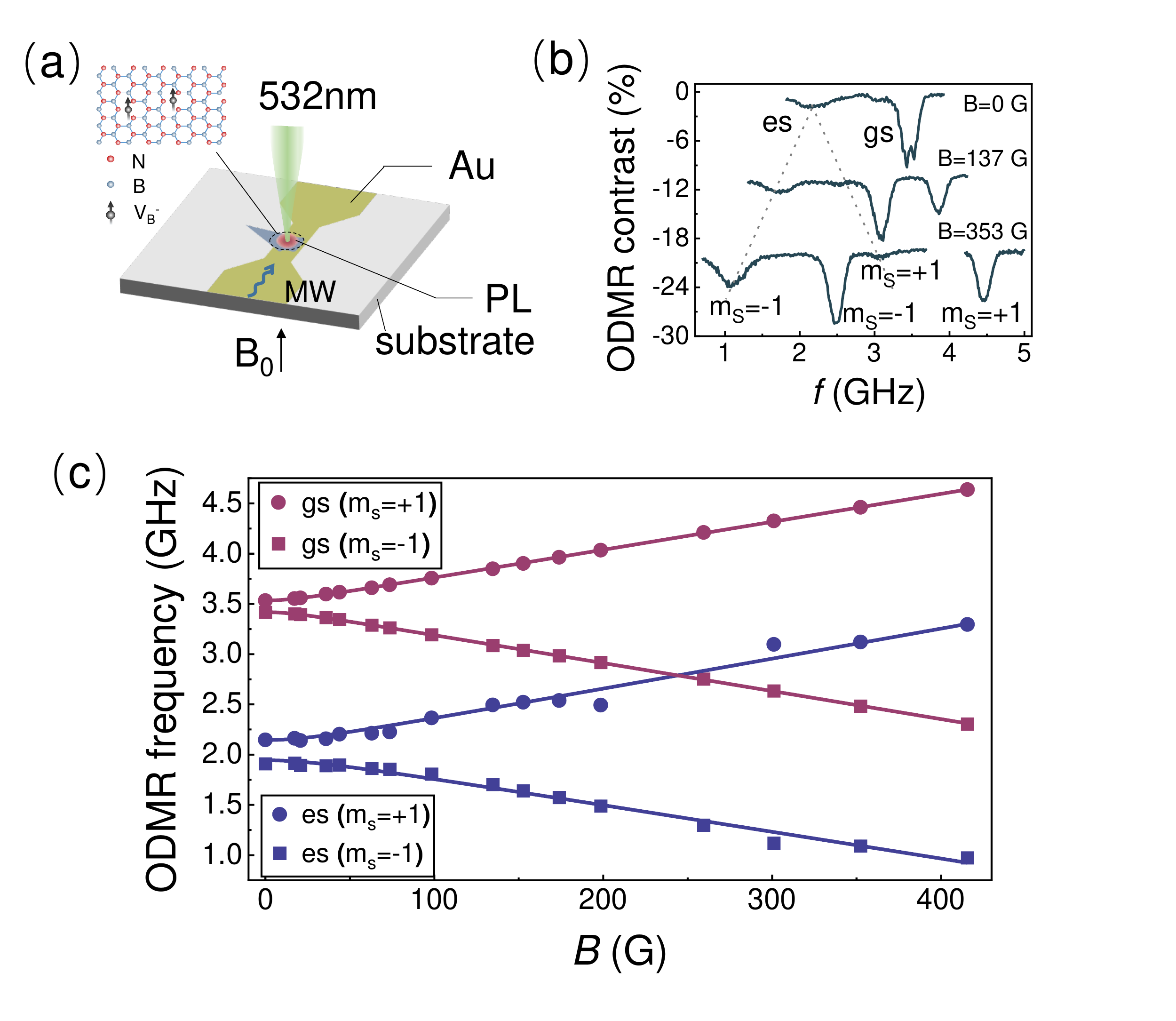}}
	\caption{ODMR of $ V_B^- $ defects at room temperature of 300K.
	(a) Experimental schematic. A hBN single crystal was transferred to the surface of gold-film coplanar microwave waveguide  after ion implantation, and microwave delivered by the waveguide was used to manipulate spins of $ V_B^- $   defects. 532 nm laser was focused on hBN sample and PL was collected by confocal system. (b) ODMR spectra of $ V_B^- $ defects at magnetic field B = 0 G, B = 137 G and B = 353 G respectively. Both ODMR in the ground state(gs) and the excited state(es) are observed (upper panel), and the broad peak in the excited state splits to two peaks at B = 137 G (middle panel) and B = 353 G (lower panel). (c) ODMR frequency versus magnetic field applied along $ c $ axis of hBN. Fitting (solid lines) using equation (1) yields parameters $ D_{es} $ = 2.06±0.04 GHz, $ E_{es} $ = 93.1± 23.3 MHz, $ g_{es} $ = 2.04±0.05 .
	}\label{fig1}
\end{figure}

We studied an ensemble of negatively charged boron vacancies created in high-purity hexagonal boron nitride single crystals, by implanting 2.5 keV helium ions with a dose of $ 2 {\rm E13/cm^2} $. These centers can be addressed via our home-built confocal photoluminescence microscope using 532 nm laser excitation. ODMR spectra of $ V_B^- $ defects were realized by applying microwaves. Tape-exfoliated hBN flakes with a thickness of tens of nanometers were directly transferred to a gold-film coplanar microwave waveguide surface. The coplanar microwave waveguide was fabricated by photolithography on an insulating substrate surface to achieve spin control and plasmonic-enhanced high ODMR contrast (Fig.\ref{fig1}a) \cite{gao2021high}. Here the microwave drive field can make Rabi frequency up to 40 MHz. ODMR spectra were performed by sweeping the frequency under continuous optical illumination and measuring $ I_{\rm {PL}} $. When the microwave driving field is not resonant with a spin transition, the laser polarizes the $ V_B^- $ defects into the $ m_S=0 $ spin state, resulting in a maximum value of $ I_{\rm {PL}} $. In contrast, when microwave driving field is resonant with a transition between the $ m_S = 0 $ and $ m_S = \pm1 $ levels, the $ I_{\rm {PL}} $ is reduced due to the lower photoluminescence rate in $ m_S = \pm1 $ \cite{gottscholl2020initialization}. If the optical transition is spin conserving, spin rotations in both the ground state and the excited state can be detected \cite{fuchs2008excited}. Fig.\ref{fig1}b shows a typical continuous wave (cw) ODMR measurement at B = 0 G. In addition to the well-known two resonant peaks, corresponding to the transitions between the ground state $ m_S = 0 $ and $ m_S = \pm1 $, which are detected near 3.48 GHz, there is also a relatively broad resonant peak near 2.1 GHz, and the broad peak splits to two peaks when external magnetic field is applied. We tentatively attribute them to spin transitions in the orbital excited state.

\begin{figure}
	\centering
	{\includegraphics[width=1.0\columnwidth]{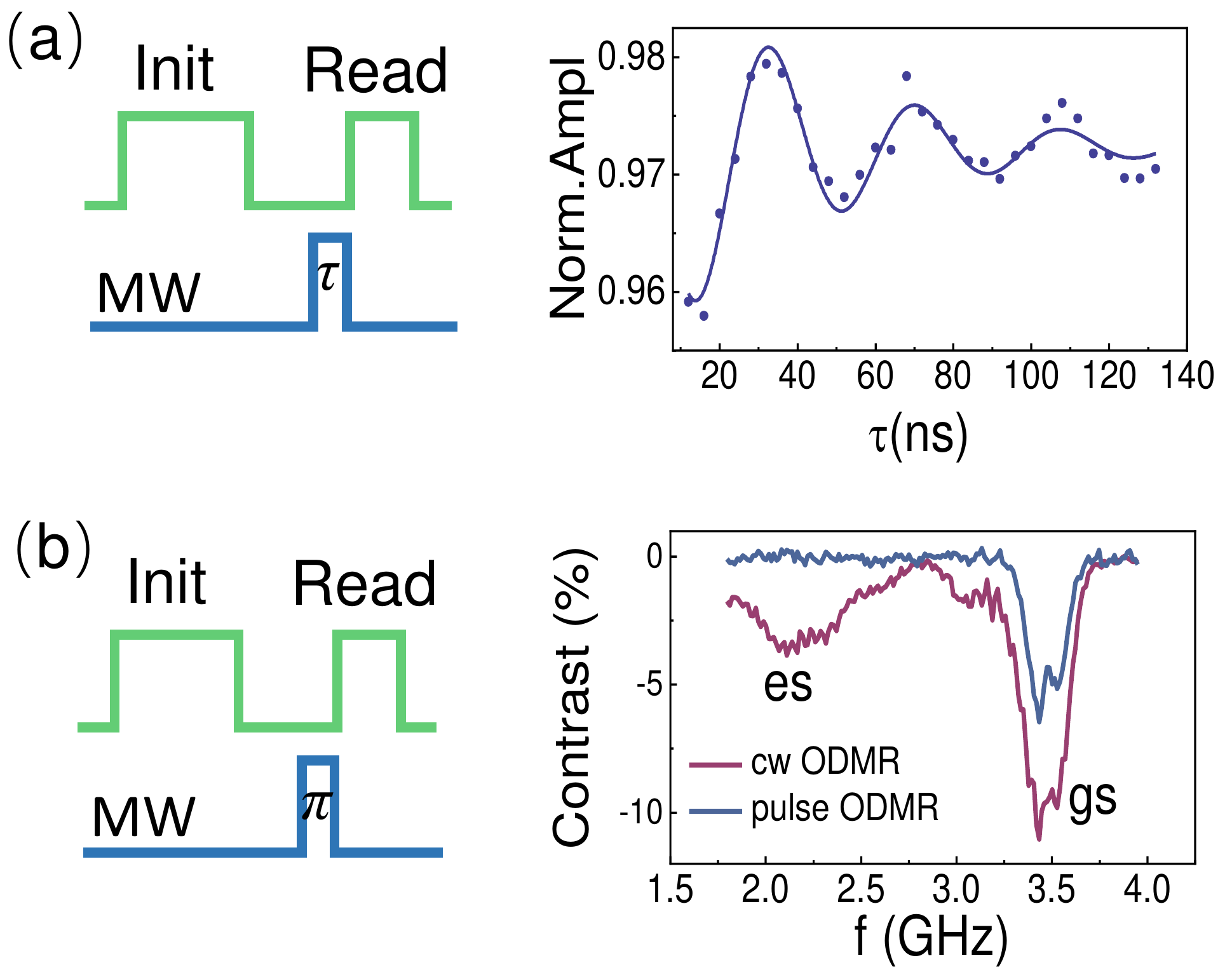}}
	\caption{Verification of ODMR in the excited state.
		(a) Pulse sequence used to measure Rabi oscillation (left) and Rabi oscillation in the ground state of $ V_B^- $ defects (right). (b) Pulse sequence used to measure pulsed ODMR, where MW was applied while laser was off to manipulate spins only in the ground state(left). Pulsed ODMR and cw ODMR of $ V_B^- $ defects (right).
	}\label{fig2}
\end{figure}

To verify that the peaks observed in cw ODMR near 2.1 GHz come from spin transitions in the excited state, we first studied the dependence of resonant frequencies on the magnitude of the applied magnetic field \cite{gottscholl2020initialization,fuchs2008excited,neumann2009excited}.  The variation of resonant frequency in cw ODMR with magnetic field applied parallel to hexagonal $ c $ axis of hBN is shown in Fig.\ref{fig1}c. Resonant frequency of spin transitions between $ m_S=0 $ and $ m_S=-1(+1) $ in the ground state starts from 3.48 GHz, and tracks to lower(higher) frequency as B increases. The excited state spin transitions have different zero field splitting of about 2.1 GHz, and
shift along the path nearly parallel to each of the ground state line, indicating the g factor of the excited state is similar to the value for the ground state \cite{gottscholl2020initialization}.
To explain the observed resonant frequencies and their variation with magnetic field, we use the excited state spin Hamiltonian:
\begin{equation}\label{Gibbs}
	H=D_{es}(S_Z^2 - S(S + 1)) + {\mu_B}{g_{es}}{\vec{B}}\cdot{\vec{S}} + E_{es} (S_X^2 - S_Y^2 ) - A_{es}\vec{S}\cdot\vec{I},
\end{equation}
where $ D_{es} $ is the excited state longitudinal (zero field) splitting, and $ E_{es} $ is the coefficient of strain-induced transverse anisotropy,  $ \vec{S} $ is the $ V_B^- $ spin-1 in the excited state with $ S_X $, $ S_Y $, $ S_Z $
components, $ g_{es} $ is the Landé factor in the excited state, $ \mu_B $ is the Bohr magneton, $ A_{es} $ is the hyperfine splitting constant. According to equation (1), resonant frequencies in the excited state vary with magnetic field B as:
\begin{equation}\label{Gibbs}
	 {\nu_{es}}_{1,2}={\nu_{es}}_0\pm\sqrt{{E_{es}}^2+(g_{es}\mu_BB)^2}/h,
 \end{equation}
where $ \nu_{es}=D_{es}/h $, fitting the data in Fig.\ref{fig1}c yields $ D_{es}=2.06 \pm 0.04 {\rm GHz} $, $ E_{es}=93.1 \pm 23.3 {\rm MHz} $, $ g_{es}=2.04 \pm 0.05 $ , indicating that the angular momentum doesn’t play an important role in the ES at room temperature \cite{gottscholl2020initialization,fuchs2008excited,neumann2009excited}. Further verification of spin transitions in the ES was conducted by using the pulsed ODMR method. Before pulsed ODMR measurement, Rabi oscillation was measured to obtain MW $ \pi $-pulse duration (Fig.\ref{fig2}a). In pulsed ODMR measurement, $ V_B^- $ defects were initialized to $ m_S = 0 $ state by optical pumping for 3 us, following a 1 us wait time to ensure spins had relaxed to the ground state. Then a $ \pi $-pulse of 13 ns was applied to flip spin to $ m_S = -1 $ state. After that, another laser pulse was applied to read out $ I_{\rm {PL}} $. One can find that the broad peaks are not visible in pulsed ODMR spectrum while the GS transitions remain observable (Fig.\ref{fig2}b), which confirms that the broad peaks  near 2.1 GHz do come from the excited state.

\begin{figure}
	\centering
	{\includegraphics[width=1.0\columnwidth]{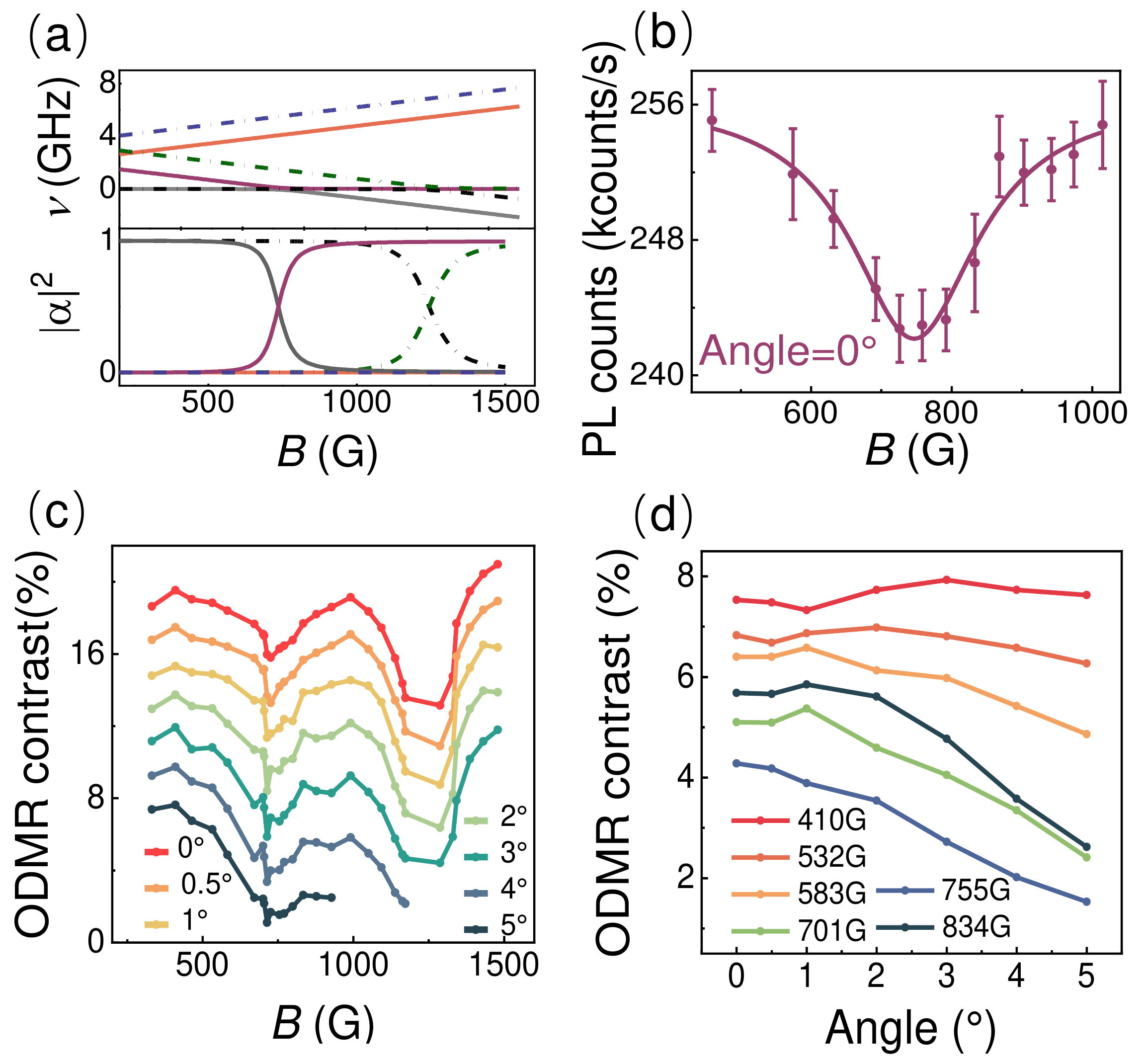}}
	\caption{Controlled level mixing by means of magnetic-field alignment.
		(a) Upper panel: calculated excited-state splitting (solid lines) and ground-state splitting (dashed lines) at $ \theta= 3^{\circ} $, where $ \theta $ is the angle between applied magnetic field and $ c $ axis of hBN. Lower panel: overlap $ |\alpha|^2 $ of each spin level with $ |0\rangle_Z $ at the angle. (b) $ I_{\rm {PL}} $ of $ V_B^- $ at different magnetic field with $ \theta=0^{\circ} $. (c) Normalized ODMR contrast versus magnetic field with different angle of misalignment. (d) Normalized ODMR contrast versus misalignment angle at different magnetic field.
	}\label{fig3}
\end{figure}

According to Hamiltonian in equation (1), level anti-crossing(LAC) is predicted around 750 G and 1240 G, corresponding to the excited state and the ground state respectively. Calculated excited state and ground state spin levels as a function of the projection of $ B $ along  $ c $ axis of hBN are showed in upper panel of Fig.\ref{fig3}a. Here the angle $ \theta $ between $ B $ and $ c $ axis of hBN is $  3^{\circ}$. The lower panel shows variation of $ |\alpha|^2 $ with B, where $ |\alpha| $ is overlap of $ |0\rangle_Z $ with each spin level, $ |m_S\rangle=\alpha|0\rangle_Z+\beta|-1\rangle_Z+\gamma|+1\rangle_Z $, $ |\beta|^2 $ and $ |\gamma|^2 $ are other overlaps respectively, and the subscript $ Z $ denotes the $ c $ basis \cite{epstein2005anisotropic,jacques2009dynamic}. Variation of $ |\alpha|^2 $ with $ B $ demonstrates spin mixing in the excited state and the ground state. $ I_{\rm {PL}} $ of $ V_B^- $ defects versus with $ B $ is plotted in Fig.\ref{fig3}b, where a peak around 750 G is observed in $ I_{\rm {PL}} $.

To further investigate the peak around the LAC, variance of ODMR contrast was measured by varying the angle $ \theta $ of magnetic field (Fig.\ref{fig3}c). Two broad peaks appear around 750G and 1240G corresponding to LAC of the excited state and the ground state respectively. This can be understood by noting the spin mixing between different spin states increases with angle around LAC, which results in a reduction of the ODMR contrast (Fig.\ref{fig3}a). The larger the tilted angle is, the more spin state mixes and the contrast reduces (Fig.\ref{fig3}c,d). However, for magnetic field B alignes along c axis ($ \theta = 0^{\circ} $), one can still observe the reduction of the ODMR contrast as well as the PL counts around LAC (Fig.\ref{fig3}b,c). We attribute this to the residual spin mixing induced by an effective tilted magnetic field due to hyperfine interaction to the surrounding nuclear spins \cite{epstein2005anisotropic,he1993paramagnetic}.

\begin{figure}
	\centering
	{\includegraphics[width=1.0\columnwidth]{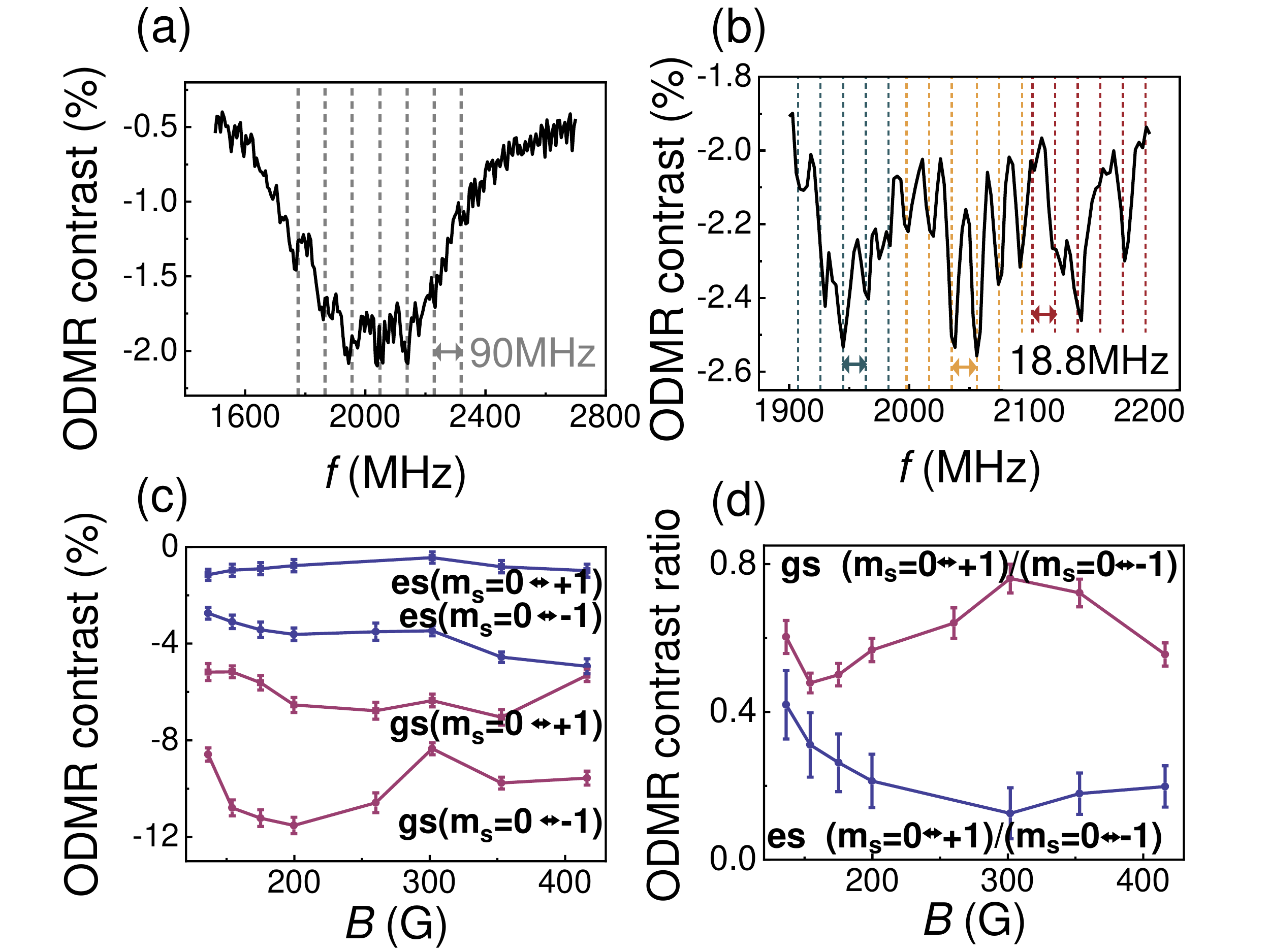}}
	\caption{Hyperfine structure in the excited-state ODMR spectrum.
		(a) ODMR spectrum of $ V_B^- $ in the excited state at B = 0 G. (b) Fine frequncy scanning of ODMR near 2.1 GHz in the excited state. (c) Variation of ODMR contrast of different spin transitons with magnitude of magnetic field. (d) Ratio of ODMR contrast in the transition between $ m_S=0 $ and $ m_S=+1 $ to transition between $ m_S=0 $ and $ m_S=-1 $ in the excited state and ground state.
	}\label{fig4}
\end{figure}

To investigate the hyperfine interaction in the ES, a fine frequency scanning of the ES peaks was performed. Fig.\ref{fig4} reveals that the resonance in ES is actually composed of a few peaks. In comparison with the hyperfine splitting of the GS spin levels by three equivalent nitrogen nuclear spins \cite{gottscholl2020initialization}, the ES spin levels also show a same pattern of seven peaks but with larger interaction $\approx$ 90 MHz ((Fig.\ref{fig4}a). This reflects the larger density of the ES wavefunction near the N nucleus. In addition to coupling to the N nucleus, one can aslo observe a few more dips inside each peak with a frequency distance around 18.8 MHz ((Fig.\ref{fig4}b). This is probably due to the hyperfine interaction to the Boron nucleus that is a bit further away than the Nitrogen.

As can be seen in Fig.\ref{fig1}b, ODMR contrast in transitions of $ m_S=0 $ to $ m_S=+1 $ is much lower than transitions of $ m_S=0 $ to $ m_S=-1 $ in the excited state. This relation is observed in different magnetic fields (Fig.\ref{fig4}c,d). We tentatively attribute it to the different lifetime of $ m_S=-1 $ and $ m_S=+1 $ state in the excited state. Since our microwave driving rate is much lower than the decay rate of ES state, the microwave induced population transfer from  $ m_S=0 $ to other two states is thus linear proportional the lifetime of each spin state. The results indicate more than two times difference of the lifetime between $ m_s=-1 $ and $ m_s=+1 $ state in ES.

In summary, we have presented a study of the excited state of $ V_B^- $ centers in hBN using cw and pulsed ODMR techniques. Spin transitions in the excited state at room temperature were observed in cw ODMR and verified by pulsed ODMR experiment. Besides, variation in cw ODMR contrast around LAC revealed the coupling with other spins. This work provides significant insights for studying the structure of $ V_B^- $ in the excited state, which is important to the implementation of dynamic nuclear polarization and ultrahigh fidelity quantum gates using the excited states.

The fabrication was performed at the USTC Center for Micro and Nanoscale Research and Fabrication, and the authors particularly thank X.W. Wang, C.L. Xin and H.F. Zuo for their assistance in ion implantation, UV lithography and deposition prosess. This work was supported by the National Key R and D Program of China (Grants No. 2018YFA0306600, 2017YFA0305000), the National Natural Science Foundation of China (Grants No. 81788101, 11775209, 12104447, 2030040389), the CAS (Grants No. XDC07000000, GJJSTD20200001, QYZDYSSW-SLH004), the Anhui Initiative in Quantum Information Technologies (Grant No. AHY050000), the Fundamental Research Funds for the Central Universities, and USTC Research Funds of the Double First-Class Initiative (Grant No. YD2340002004), China Postdoctoral Science Foundation (Grant No. 2020M671858).

\bibliographystyle{naturemag}

\bibliography{hBN.bib}

\end{document}